\documentclass[twocolumn]{svjour3}

\usepackage{graphicx}
\usepackage{amsmath}
\usepackage{array} 
\usepackage{dcolumn}
\usepackage{bm}
\usepackage{color} 
\usepackage{ulem}
\usepackage{widetext}

\newcommand{\nwip}[2]{\langle #1|#2\rangle} 

\begin{document}

\title{Detection and Measurement of Micrometeoroids with LISA Pathfinder}

\author{ J.~I.~Thorpe \and C.~Parvini \and J.~M.~Trigo-Rodr\'{i}guez}

\institute{J.~I.~Thorpe \and C.~Parvini \at Gravitational Astrophysics Laboratory, NASA Goddard Space Flight Center, Greenbelt, MD 20771, USA\email{james.i.thorpe@nasa.gov}
\and
C.~Parvini \at Department of Aerospace Engineering, The George Washington University, Washington, DC 20052, USA
\and
J.~M.~Trigo-Rodr\'{i}guez \at Meteorites, Minor Bodies and Planetary Sciences Group, Institute of Space Sciences (CSIC-IEEC), Campus UAB Bellaterra, Carrer de Can Magrans, s/n 08193 Cerdanyola del Vall\'{e}s (Barcelona), Spain.
}

\date{\today}

\maketitle

\abstract{
The Solar System contains a population of dust and small particles originating from asteroids, comets, and other bodies.  These particles have been studied using a number of techniques ranging from in-situ satellite detectors to analysis of lunar microcraters to ground-based observations of zodiacal light. In this paper, we describe an approach for using the LISA Pathfinder (LPF) mission as an instrument to detect and characterize the dynamics of dust particles in the vicinity of Earth-Sun L1. Launching in late 2015, LPF is a dedicated technology demonstrator mission that will validate several key technologies for a future space-based gravitational-wave observatory. The primary science instrument aboard LPF is a precision accelerometer which we show will be capable of sensing discrete momentum impulses as small as $4\times 10^{-8}\,\textrm{N}\cdot\textrm{s}$. We then estimate the rate of such impulses resulting from impacts of micrometeoroids based on standard models of the micrometeoroid environment in the inner solar system. We find that LPF may detect dozens to hundreds of individual events corresponding to impacts of particles with masses $> 10^{-9}\,$g during LPF's roughly six-month science operations phase in a $5\times 10^5\,\textrm{km}$ by $8\times 10^5\,\textrm{km}$ Lissajous orbit around L1. In addition, we estimate the ability of LPF to characterize individual impacts by measuring quantities such as total momentum transferred, direction of impact, and location of impact on the spacecraft.  Information on flux and direction provided by LPF may provide insight as to the nature and origin of the individual impact and help constrain models of the interplanetary dust complex in general. Additionally, this direct in-situ measurement of micrometeoroid impacts will be valuable to designers of future spacecraft targeting the environment around L1.
}

\section{Introduction}
\label{sec:intro}

Our current understanding of the interplanetary dust complex is informed by a number of measurement techniques including photographic and visual meteors \cite{Halliday1984,Hawkes2007,Trigo-Rodriguez2008}, radio meteors, atmospheric collections \cite{Flynn1994}, observation of zodiacal light in thermal \cite{Krick2012} and visual \cite{Durmont1980} wavelengths, in-situ penetration and ionization detectors \cite{Weiden1978,Zhang1995}, and analysis of micro-craters in lunar samples \cite{Allison1982}. Additionally, NASA's Stardust mission  \cite{Horz2006} and ESA's Rosetta mission  \cite{Rotundi2015} have made in-situ measurements of the dust environments near the comets Wild 2 and 67/Churyumov-Gerasimenko respectively. The combined picture from all of these techniques,  each with varying ability to detect and characterize various sub-populations of the dust complex, is used to constrain theoretical models that account for the sources, sinks, and dynamics of the dust complex. 

The LISA Pathfinder (LPF) mission \cite{LPF_LISAX}, is a technology demonstration mission dedicated to validating several key technologies for a future space-based observatory \cite{Bender_98,ESA_Yellow_Book} of astrophysical gravitational waves \cite{Einstein:1918aa,Bondi1959}. Led by the European Space Agency with contributions from NASA and a number of European research institutions, LPF consists of a single sciencecraft operating in a $5\times 10^5\,\textrm{km}$ by $8\times 10^{5}\,\textrm{km}$ Lissajous orbit around the first Sun-Earth Lagrange Point (L1). LPF is currently planed to launch in December of 2015 with science operations beginning in early 2016 and lasting for approximately six months with a possible extension of an additional six months. In this paper, we outline an approach for using LPF as an instrument to detect and characterize interplanetary dust in the vicinity of L1.  This technique does not require any modification to the hardware or operations of LPF and may provide an important new source of information to constrain models of the dust complex.

The outline of the paper is as follows.  In section \ref{sec:overview} we describe the working principle of the LPF instrument and its application as a dust detection instrument. In section \ref{sec:thresh} we make a simple estimate for the detection threshold of LPF as a function of transferred momentum using a simplified 1-D model. In section \ref{sec:rates} we estimate the rates of events above this threshold and the total number of likely events during the LPF mission. In section \ref{sec:pe} we estimate the precision with which LPF will measure parameters of individual impacts such as momentum transferred and direction of impact. In section \ref{sec:discuss}, we summarize our results and outline likely avenues for refining the estimates made here.

\section{Overview of Technique}
\label{sec:overview}

Inside the LPF are two inertial sensor units each containing a test mass comprised of a $46\,$mm cube of Au-Pt alloy with a mass of $1.96\,$kg. During launch and cruise, these test masses are mechanically supported by a lock mechanism, which is retracted prior to beginning science operations leaving the test masses freely-falling inside the inertial sensors with no mechanical contacts. Each inertial sensor unit measures the position and attitude of its respective test mass relative to the LPF spacecraft in all six rigid-body kinematic degrees-of-freedom (DoFs) using a capacitive sensing system. The inertial sensor unit can apply forces and torques to the test mass via electrostatic actuation. Additional information is provided by a laser interferometer that measures a reduced set of DoFs with higher precision and a star tracker which measures the three angular DoFs of the spacecraft relative to a background inertial reference frame.  A control system converts the position and attitude measurements from the two inertial sensors, the interferometer, and the star tracker into force and torque commands that are applied to the test masses and the spacecraft, the latter being actuated using a microthruster system.
 
The main purpose of the control system is to reduce external force disturbances on the test masses; low-disturbance reference masses are a key technology  for LISA-like instruments. A byproduct of this effort is an exquisite measurement of external force disturbances on the LPF spacecraft either through the residual motion between the spacecraft and the test masses or in the applied control forces and torques. One such source of external force disturbance is an impulse generated by an impact of a micrometeoroid or dust particle. In the following sections, we estimate the detection threshold and rate for such impact events.

 \section{Detection Threshold}
\label{sec:thresh}
The momentum transferred to the LPF spacecraft by an individual micrometeoroid impact will combine with stochastic disturbances on the spacecraft from a number of sources such as Solar radiation pressure, outgassing, and noise from the microthruster system.  The \textit{measured} momentum transfer will also be effected by noise in the position sensing system and force disturbances on the test masses. To simplify our estimate of the threshold at which LPF can detect and individual impact, we consider a simplified one-dimensional model in a linear DoF. We adopt the maximum-likelihood formalism, where the signal-to-noise ratio (SNR) of an event is defined using a noise-weighted inner-product:
\begin{equation}
\rho^2 = \nwip{F}{F}, \label{eq:SNR} 
\end{equation}
where $\rho$ is the SNR, $F$ is the `waveform' of the force on the spacecraft resulting from the impact and $\nwip{\ldots}{\ldots}$ denotes the noise-weighted inner product defined as:
\begin{equation}
\nwip{a}{b} \equiv 4 \mathcal{R} \int_0^{+\infty} \frac{\tilde{a}(f)\tilde{b}^*(f)}{S_n(f)} df, \label{eq:nwip}
\end{equation}
where tildes denote Fourier-domain signals and $S_n(f)$ is the one-sided power-spectral density of the equivalent force noise on the spacecraft. Given a waveform $F$ and an estimate of the noise characterized by $S_n(f)$, the SNR can be calculated using (\ref{eq:nwip}) and (\ref{eq:SNR}). An SNR threshold for detection, typically set around 5-10, can be used to determine whether an individual event is likely to be observed.

While the error budget developed for LPF contains dozens of individual effects that contribute equivalent force noise to the test masses and spacecraft, there are two effects that dominate force noise on the spacecraft in the measurement band $0.1\,\textrm{mHz}\leq f \leq 100\,\textrm{mHz}$.  The first is the noise of the microthruster system itself, which is characterized by a flat power spectral density with a level of 

\begin{equation}
S_{th}\approx 10^{-16}\,\textrm{N}^2/\textrm{Hz},\label{eq:Sth}
\end{equation}

The thruster noise dominates at low frequencies but is eclipsed above a few mHz by the equivalent force noise of the inertial sensor. The inertial sensor noise is characterized by a flat displacement amplitude spectral density at a level of $\sim2\,\textrm{nm}/\textrm{Hz}^{1/2}$. This can be converted to an equivalent force noise on the spacecraft by multiplying by the spacecraft mass ($M=422\,$kg) and taking two time derivatives.  The resulting power spectral density of this equivalent spacecraft force noise from the inertial sensor is given by

\begin{eqnarray}
S_{is} & \approx \left[(2\times 10^{-9}\,\textrm{m}/\textrm{Hz}^{1/2})\cdot M \cdot (2\pi f)^2\right]^2 \nonumber \\
& = 1.1\times 10^{-9}\cdot f^4\,\textrm{N}^2/\textrm{Hz}^5. \label{eq:Sis}
\end{eqnarray}
 
\begin{figure}[h]
\begin{center}
\includegraphics[width=8 cm]{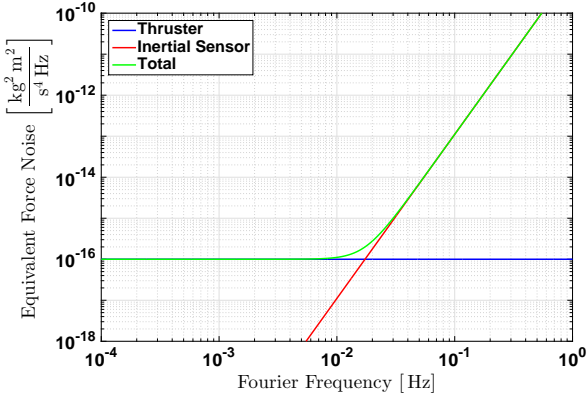}
\caption{Power spectral density of equivalent spacecraft force noise in a simplified one-dimensional model of LPF.}
\label{fig:modelNoise}
\end{center}
\end{figure}

The individual and combined components of the spacecraft equivalent force noise are shown in Figure \ref{fig:modelNoise}.

An impact by a micrometeoroid is modeled as an impulsive force occurring at time $\tau$ and imparting a total momentum $P$ over a characteristic timescale $\Delta$. This can be written in the time and frequency domains respectively as:

\begin{eqnarray}
F(t) = & \frac{P}{\Delta}\left[\Theta(t-\tau)-\Theta(t-\tau-\Delta)\right], \label{eq:F_time}\\ 
\tilde{F}(f) = & P sinc(f \Delta)e^{-2\pi i f (\tau+\Delta/2)} \label{eq:F_freq}.
\end{eqnarray}

where $\Theta(t)$ is the Heaviside step function, and $sinc(x)\equiv sin(\pi x)/\pi x$ is the normalized sine cardinal function. One would generally expect that the characteristic impulse times for micrometeoroids would be small compared to the standard sampling rate for LPF data of $1\,$Hz. In this limit,

\begin{equation}
\lim_{\Delta\to 0} \tilde{F}(f)\approx P e^{-2\pi i f \tau}.
\label{eq:Fwave}
\end{equation}

This provides us with a waveform parametrized by two parameters, total momentum transferred $P$ and impact time $\tau$.  Using the expression for $\tilde{F}(f)$ in (\ref{eq:Fwave}) and a generic total equivalent force noise of the form $S_n(f)\equiv S_{0} + S_{4}f^4$, the formula for SNR in (\ref{eq:SNR}) can be used to determine the SNR as a function of $P$:

\begin{eqnarray}
\rho = & P/P_c, \nonumber \\
P_c \equiv & \frac{1}{\sqrt{2\pi}}\left(4 S_4 S_0^3\right)^{1/8}\label{eq:SNRp}
\end{eqnarray} 
where $P_c$ is the characteristic threshold momentum. For the case described above of $S_0 = 10^{-16}\,\textrm{N}^{2}/\textrm{Hz}$ and $S_4 = 1.1\times 10^{-9}\,\textrm{N}^2/\textrm{Hz}^5$, the characteristic momentum $P_c \approx 3.6\times 10^{-8}\,\textrm{N}\cdot\textrm{s}$. For an event to be detected with $\rho\ge 8$, the total momentum must be $P\ge 8 P_c \approx 2.9\times 10^{-7}\,\textrm{N-s}$.

 \section{Rate Estimate}
\label{sec:rates} 
The population of dust and micrometeoroids in the inner solar system is derived primarily from the collisional processing of asteroids and comets. The particles making up this population vary in mass, size, composition, and orbit and combine to form a dust complex with a complex morphology \cite{2004come.book..677S}. The most commonly-used model of the population is the model of Gr\"{u}n, et al. \cite{Grun1985}, which estimates the cumulative flux of micrometeoroids in the inner solar system as a superposition of three distinct populations:

\begin{widetext}
\begin{eqnarray}
\Phi_1(m) & =& \left(2.2\times10^3 m^{0.306}+15.0\right)^{-4.38},\: 10^{-9}\,\textrm{g} < m < 10^{0}\,\textrm{g}, \nonumber \\
\Phi_2(m)& = &1.3\times 10^{-9}\left(m+10^{11}m^2+10^{27}m^4\right)^{-0.36},\: 10^{-14}\,\textrm{g} < m < 10^{-9}\,\textrm{g}, \nonumber \\
\Phi_3(m) &= &1.3\times 10^{-16}\left(m+10^{6}m^2\right)^{-0.85},\: 10^{-18}\,\textrm{g} < m < 10^{-14}\,\textrm{g}, \nonumber \\
\Phi(m) & = &3.15576\times10^7\left[\Phi_1(m) + \Phi_2(m) + \Phi_3(m)\right].
\label{eq:grun}
\end{eqnarray}
\end{widetext}

\begin{figure}[h]
\begin{center}
\includegraphics[width=8 cm]{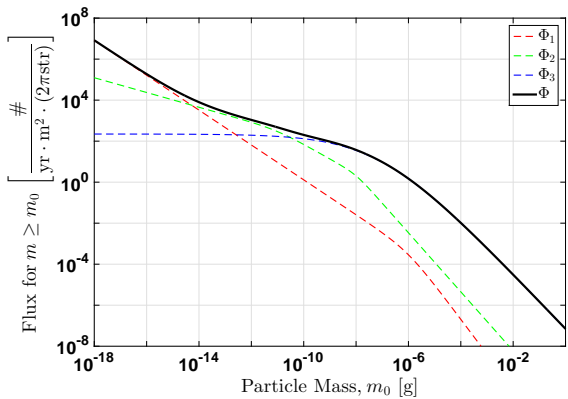}
\caption{Flux of micrometeoroids with mass $m\ge m_0$ in the inner Solar system from the Gr\"{u}n, et al. \cite{Grun1985} model. The dashed lines show the individual components and the solid line shows the total flux as described in (\ref{eq:grun}). }
\label{fig:grun}
\end{center}
\end{figure}

 The flux $\Phi(m)$ represents the total number of particles with mass greater than $m$ grams impacting a unit area from a single hemisphere ($2\pi$ steradians) in a single year. Figure \ref{fig:grun} contains a plot of this model. Much of the literature on micrometeoroid flux addresses issues related to suppression or enhancement of various regions in this power-law due to effects of the Earth and Moon, but for LPF's orbit around Sun-Earth L1, the unmodified Gr\"{u}n model is most appropriate.
 
For detection with LPF, the mass distribution of the micrometeoroid population is just part of the required information. The transferred momentum additionally depends on the relative velocity between the micrometeoroid and the spacecraft. As a zeroth-order estimate, we assume that the characteristic velocity of impacts is equivalent to the orbital velocity of LPF around the Sun, or roughly $30\,\textrm{km}/\textrm{s}$.  Under the admittedly crude assumption that all impacts occur with this velocity, we can compute the rate of LPF detections as
 \begin{equation}
R \approx A\Phi\left(\frac{P_c\rho_0}{\bar{v}}\right),
\label{eq:rate}
\end{equation}

where $\Phi(m)$ is the Gr\"{u}n model flux from (\ref{eq:grun}), $P_c$ is the characteristic momentum from (\ref{eq:SNRp}), $\rho_0$ is the SNR threshold for detection, $\bar{v}$ is the characteristic velocity, and $A$ is the cross-sectional area of the spacecraft.  For $\rho_0=8$ and $A=3\,\textrm{m}^2$, (\ref{eq:rate}) gives $R\approx 1.2\times 10^2\,\textrm{yr}^{-1}$ or about 60 events over the 180-day baseline LPF science mission.

\begin{figure}[h]
\begin{center}
\includegraphics[width=8 cm]{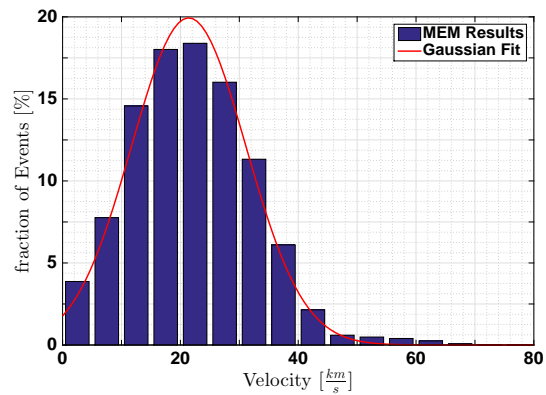}
\caption{Probability distribution of impact velocities for LPF micrometeoroid collisions during science operations. Histogram in blue are estimates from the NASA Meteoroid Engineering Model and a representative ephemeris for LPF. The fit in red is a best-fit normal distribution. }
\label{fig:Vdist}
\end{center}
\end{figure}

 To improve upon this rate estimate, we employed the NASA Meteoroid Engineering Model (MEM) \cite{McNamara:2004aa}, which was developed to help spacecraft designers and mission planners assess micrometeoroid risk from impacts. To quantify such risk, it is crucial to  know the total number of impacts, their mean velocity, and the size of the biggest meteoroid that the spacecraft is likely to encounter. The MEM takes as input a state vector describing the orbit of a spacecraft and returns the total flux and velocity distribution of impacts for particles having a mass $m\ge 10^{-6}\,$g.  Because the MEM concentrates on these higher-mass events, which are more rare, it is not by itself useful in determining the number of the events likely to be detected by LPF. However, we can use the MEM to provide an improved estimate of the distribution of impact velocities under the assumption that this distribution is roughly constant as we extend to lower masses. Using a representative state vector for the LPF science orbit provided by ESA, the MEM was used to produce the velocity distribution in Figure \ref{fig:Vdist}.  The MEM results are reasonably well fit by a normal distribution with a mean velocity of $\bar{v}=21.4\,\textrm{km}/\textrm{s}$ and a standard deviation of $\sigma_v = 9.7\,\textrm{km}/\textrm{s}$.

\begin{figure}[h]
\begin{center}
\includegraphics[width=8 cm]{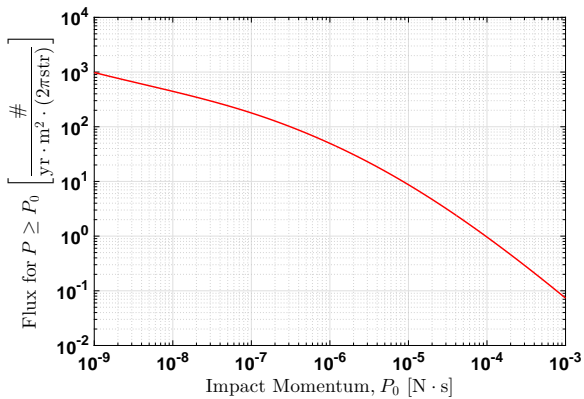}
\caption{Flux of LPF particle impacts with transferred momenta $P\ge P_0$ estimated using the Gr\"{u}n model mass flux in Figure \ref{fig:grun} and the fit to the MEM-derived velocity distribution in Figure \ref{fig:Vdist}. }
\label{fig:Pdist}
\end{center}
\end{figure}

Combining the mass flux from the Gr\"{u}n model and the velocity distribution from Figure \ref{fig:Vdist} produces the impact flux as a function of particle momentum in Figure \ref{fig:Pdist}. Note that for a threshold momentum of $8P_c \approx 1.7\times 10^{-7}\,\textrm{N}\cdot\textrm{s}$, the predicted event rate assuming a $3\,\textrm{m}^2$ area is approximately $3\times 10^2$ events per year, about $2.5$ times higher than the simple prediction in (\ref{eq:rate}). This increase is due to detections of the high-velocity tail of the lighter particles, which are more numerous. 
 
An improved treatment of the event rate would include a geometrical model for the LPF spacecraft and its orientation as well as any correlations between impact mass and impact velocity.  It may also be possible to forecast the impact rates from asteroidal dust trails or cometary meteoroid showers that intersect LPF's orbit.

 \section{Parameter Estimation}
 \label{sec:pe}
 A key feature of LPF as a micrometeoroid instrument is the ability to characterize in addition to simply detecting micrometeoroid impacts.  For example, each LPF test mass will measure the transferred linear momentum from the impact in three orthogonal directions.  The error of in the estimate of each of these momentum components will be $\sim\rho^{-1}$ where $\rho$ is the SNR defined in (\ref{eq:SNRp}).  Under the assumption that the error in each momentum component is independent, the two angles describing the impact direction will be measured with errors of roughly
 \begin{equation}
\sigma_\theta \sim \sigma_\phi \sim \sqrt{3}\rho^{-1}.
\label{eq:errAng}
\end{equation}
For example, an event with $\rho = 10$ will have typical errors in the impact angles of $0.17\,\textrm{rad}\approx 10^{\circ}$. Knowledge of the direction of the impact combined with an ephemeris for LPF will allow reconstruction of the impactor's orbit, a key piece of information for distinguishing different populations of micrometeoroids. For example, it may be possible to measure an excess of impacts coming from a particular direction that could be associated with a known comet. This would allow the velocity of the impact to be inferred from the ephemerides of LPF and the comet which would in turn allow an estimate of the mass of each impacting particle from the measured momentum transfer.

LPF will also measure three components of the angular momentum transferred to the spacecraft. To estimate the noise floor for angular degrees of freedom, we follow a similar procedure to the analysis in section \ref{sec:thresh}.  The angular sensing noise of the LPF inertial sensor units is characterized by a flat spectrum with an amplitude spectral density of $\sim 200\,\textrm{nrad}/\textrm{Hz}^{1/2}$. This can be converted into an equivalent torque noise by multiplying by the spacecraft moment of inertia $\sim 200\,\textrm{kg}\cdot\textrm{m}^2$ and taking two time derivatives. The resulting power spectral density of the equivalent torque noise is

\begin{eqnarray}
S_{Nis} & \approx \left[(2\times 10^{-7}\,\textrm{rad}/\textrm{Hz}^{1/2})\cdot I \cdot (2\pi f)^2\right]^2 \nonumber \\
& = 2.5\times 10^{-6}\cdot f^4\,\textrm{N}^2\cdot\textrm{m}^2/\textrm{Hz}^5. \label{eq:SNis}
\end{eqnarray}
 
\begin{figure}[h]
\begin{center}
\includegraphics[width=8 cm]{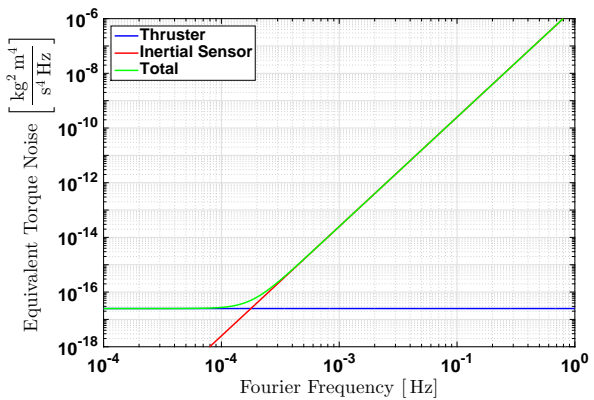}
\caption{Power spectral density of equivalent spacecraft torque noise in a simplified model of LPF.}
\label{fig:modelNoiseTorque}
\end{center}
\end{figure}

The equivalent torque noise of the micropropulsion system can be estimated by multiplying the force noise by a characteristic `lever-arm' corresponding to the perpendicular distance between the spacecraft center of mass and a line along the thrust vector passing through the thruster mounting point. A rough estimate of this value is $0.5\,$m based on the physical dimensions of the LPF spacecraft. This gives an equivalent torque noise of $S_{Nth}\approx 2.5\times 10^{-17}\,\textrm{N}^2\cdot\textrm{m}^2/\textrm{Hz}^2$. Figure \ref{fig:modelNoiseTorque} shows the power spectral density of the total equivalent torque noise for this simplified model.

The characteristic angular momentum for LPF can then be computed using (\ref{eq:SNRp}) with $S_0 = 2.5\times 10^{-17}\,\textrm{N}^2\cdot\textrm{m}^2/\textrm{Hz}^2$ and $S_4 = 2.5\times 10^{-6}\cdot f^4\,\textrm{N}^2\cdot\textrm{m}^2/\textrm{Hz}^5$. The result is $L_c = 5.6\times 10^{-8}\,\textrm{N}\cdot\textrm{m}\cdot\textrm{s}$. The typical errors on impact location can then be estimated as

 \begin{equation}
\sigma_r= \rho^{-1}\frac{L_c}{P_c} \approx \rho^{-1} 1.6\,\textrm{m}
\label{eq:errPos}
\end{equation}

A $\rho = 10$ event would have typical localization errors of $\sim16\,\textrm{cm}$. A more detailed reconstruction of the impact location would include a mechanical model of the spacecraft and might improve the impact localization somewhat. While the impact location is not as useful a measurement for studying the micrometeoroid population as the total momentum or impact direction, it may be of interest to LPF's operations team.

 \section{Discussion}
 \label{sec:discuss} 
The next logical step in this research effort is to refine the estimates made here, taking advantage of the detailed information available about the LPF spacecraft and additional nuances of the known or modeled micrometeoroid environment. Examples of improvements in the LPF model might include proper modeling of the noise and noise correlations in the inertial sensor and thruster subsystems, including the second inertial sensor as a quasi-independent sensor, and an improved rate estimate including a geometrical model of the LPF spacecraft and its orientation during the science operations. 

Modifications to the micrometeoroid flux model might include recent radar data ( \cite{Janches2015} and references therein) that indicate a bimodal distribution of velocities in which the fraction of high-velocity meteoroids peaking at $50-60\,$km/s. If such a high-velocity population exists in the vicinity of L1, it would extend LPF's detection capability to lower mass particles and further boost the detection rate. Similarly, micrometeoroids coming from high-eccentricity cometary orbits collide at higher velocity than the asteroidal one, so the momentum transferred will be higher and the threshold for mass detection will be lower. We think that such capability to compute the flux from the cometary source and its possible fine-structure associated with the disintegration of periodic comets may also be of interest. 

In addition to improving estimates of detection thresholds, event rates, and parameter estimation capabilities, an analysis pipeline could be developed that would detect and characterize events from a stream of LPF science data.  Techniques that have already been developed for the analysis of data from ground- and space-based gravitational wave detectors would likely be applicable to this problem.

\section{Conclusion}
\label{sec:conclude}
We have outlined a concept for using the LPF spacecraft as an instrument for studying the micrometeoroid environment in the vicinity of Earth-Sun L1 and made rough estimates of the detection threshold, event rates, and errors with which physical parameters are likely to be measured.These estimates suggest that LPF should be capable of detecting dozens to hundreds of individual events during its operational lifetime and will measure their individual momenta, impact directions, and impact locations with interesting precision. 

LPF represents a unique and novel opportunity to make in-situ measurements of the micrometeoroid environment in the vicinity of L1, a region for which no in-situ measurements exist. Such measurements can be used to constrain models of the micrometeoroid population in the inner solar system and will complement existing measurements using other techniques.  Our proposed method requires no modification to LPF's hardware or operations, it simply requires a dedicated analysis of the data that will be collected as part of normal science operations.

\section*{Acknowledgements}
The authors would like to acknowledge Ian Harrison for providing the representative LPF ephemeris file used to estimate the distribution of impact velocities in section \ref{sec:thresh}. JMTR's research was supported by the Spanish Ministry of Science and Innovation (project: AYA2011-26522). CP's research was supported by the 2015 NASA Goddard Space Flight Center Summer Internship Program. Copyright (c) 2015 United States Government as represented by the Administrator of the National Aeronautics and Space Administration. No copyright is claimed in the United States under Title 17, U.S. Code. All other rights reserved.

\bibliographystyle{spmpsci}
\bibliography{Bibliography.bib}
\end{document}